\input{aipcheck}

\documentclass[
    ,final            
  ]
  {aipproc}

\layoutstyle{8x11single}

\begin{document}

\title{Diffractive pion production at COMPASS -- First results on 3$\pi$ final
states - neutral mode}


\author{Frank Nerling for the COMPASS Collaboration}{
  address={Physikalisches Institut, Universit\"at Freiburg,\\ Hermann-Herder-Str. 3, D-79104 Freiburg, Germany}
}

\begin{abstract}
The COMPASS experiment at CERN is designed for light hadron spectroscopy with emphasis
on the detection of new states, in particular the search for exotic states and glue-balls.
After a short pilot run in 2004 (190 GeV/c negative pion beam, lead target) showing significant 
production strength for an exotic $J^{PC}=1^{-+}$ state at 1.66\,GeV/${\rm c^2}$, we have collected data 
with a 190 GeV/c negative charged hadron beam on a proton (liquid hydrogen) and nuclear targets in 2008 and 2009. 
The spectrometer features good coverage by electromagnetic calorimetry, and our data provide excellent 
opportunity for simultaneous observation of new states in two different decay modes in the 
same experiment. The diffractively produced $(3\pi)^{-}$ system for example can be studied in both 
modes $\pi^{-}p \rightarrow \pi^{-}\pi^{+}\pi^{-}p$ and $\pi^{-}~p \rightarrow \pi^{-}\pi^{0}\pi^{0}~p$.
Charged and neutral mode rely on completely different parts of the spectrometer. Observing a new
state in both modes provides important cross-check. 
First results of a preliminary PWA performed on the 2008 data are presented. 
\end{abstract}

\maketitle


\section{Introduction}
The COMPASS fixed target experiment~\cite{Compass:1996} at CERN SPS is dedicated to the study of nucleon spin 
structure and hadron spectroscopy, addressing the question of how nucleons and hadrons in general are built 
up from quarks and gluons~\cite{CompassHadronProposal:2007}. 
%
The COMPASS Collaboration has already collected data scattering a polarised 160\,GeV/c muon beam on polarised 
deuteron ($^{6}$LiD) and proton (NH$_{3}$) targets during the years 2002-2004 and 2006-2007. 
The gluon contribution to the nucleon spin is one example of physics determined from these data. 
During a second phase dedicated to physics with hadron beams, we have collected unprecedented statistics of data 
with 190 GeV/c charged hadron beams on a proton and nuclear targets in 2008 and 2009. The feasibility of our 
apparatus for light mesons spectroscopy has been studied in a short pilot run in 2004 (190 GeV/c negative pion beam, 
lead target). 
%
Based on the few days diffractive pion data included in the 2004 run, pion dissociation into 
$\pi^{-}\pi^{-}\pi^{+}$ final states has been analysed showing significant production strength for an exotic $J^{PC}=1^{-+}$ state at 
1.66\,GeV/${\rm c^2}$, which can be interpreted as the $\pi_1(1600)$ \cite{Alekseev:2009a}. 

The high statistics data sample taken with the improved spectrometer in 2008/09 allows us not only to complete the 
search for the $\pi_1$ but also to extent our analyses to further channels of interest (with lower cross section and 
higher masses) and further develop our PWA methods. In particular the detection of final states with both charged and neutral particles
is one of the key advantages of COMPASS as compared to previous fixed target experiments.
First preliminary results on the 2008 data for pion dissociation into 3 pion final states, neutral mode: 
$\pi^{-}\pi^{0}\pi^{0}$, are presented. The simultaneous study of both modes allows 
for important cross-check (acceptances, systematics) and independent confirmation of any new state observed in the charged mode. 
%
%
%
\section{Light meson spectroscopy -- Diffractive dissociation}
The naive Constituent Quark Model (CQM) characterises mesons as bound colour-singlet states of a quark $q$ and a anti-quark 
$\bar q$ with flavours $u,d$ and $s$ grouped into SU(3)$_{\rm flavour}$ multiplets. Their total angular momentum $J$, 
parity $P$ and charge conjugation
$C$ are given by $J=|L-S| ... |L+S|$, $P=(-1)^{L+1}$, and $C=(-1)^{L+S}$~,
where $L$ is the relative orbital angular momentum of $q$ and $\bar q$, and $S$ the total intrinsic spin $(S=0,1)$ of the $q
\bar{q}$ pair. In addition the isospin $I$ and the $G$ parity defined as $G=(-1)^{I+L+S}$ are introduced, 
also conserved in strong interactions.
Given the simplicity, the CQM is astonishingly successful in describing part of the meson properties as well as -- to a 
large extend -- the observed spectrum. 

In QCD, however, interactions between coloured quarks are described by exchange of gluons $g$ carrying colour themselves, 
resulting in the prediction of new phenomena. In particular, colour-singlet mesons are not restricted to be composed of $q\bar{q}$ pairs 
but may consist of other colour-neutral configurations, like e.g. $qq\bar{q}\bar{q}$ (tetra-quarks), $q\bar{q}g$ (hybrids) or $gg$ 
(glueballs), which are mostly discussed in this context.
Due to mixing with ordinary $q\bar{q}$ states with same quantum numbers $J^{PC}$, such configurations are extremely 
difficult to find experimentally, since it is hardly possible to disentangle the contribution of each configuration.
The experimental observation of spin-exotic mesons with quantum numbers forbidden in the CQM, like e.g. 
$J^{PC}=~0^{--}, 0^{+-}, 1^{-+}$, would thus provide a clear evidence for physics beyond the naive quark model and a fundamental confirmation 
of QCD.

The lowest-lying hybrid is expected to have $J^{PC}=1^{-+}$. Lattice-QCD simulations \cite{cmcneile:2006} and flux-tube model 
calculations \cite{fclose:1995} predict a mass between 1.7 and 2.2\,${\rm GeV/c^2}$, and a preferred decay into $b_{1}\pi$ and $f_{1}\pi$. 
Experimentally two candidates for a $1^{-+}$ hybrid have been found, $\pi_{1}(1400)$ and $\pi_{1}(1600)$, however, both are still heavily 
disputed in the community.
The $\pi_{1}(1400)$ was mainly seen in $\eta\pi$ decays, by e.g. E852~\cite{dthompson:1997}, VES\cite{gbeladidze:1993}, and Crystal 
Barrel~\cite{aabele:1998}. 
The $\pi_{1}(1600)$ was observed by both E852 and VES in the decay channels: $\rho\pi$~\cite{gadams:1998,ykhokhlov:2000}, 
$\eta'\pi$~\cite{gbeladidze:1993,eivanov:2001}, $f_{1}\pi$~\cite{jkuhn:2004,damelin:2005}, and $\omega\pi\pi$~\cite{damelin:2005,mlu:2005}. 
Especially the observations of $\pi_{1}(1600)$ into $\rho\pi$ based on analyses of $\pi^{-}\pi^{+}\pi^{-}$ final state events are controversially discussed \cite{schung:2002,adzierba:2006}. 
\begin{figure}[tp!]
  \begin{minipage}[h]{.39\textwidth}
    \begin{center}
      \includegraphics[clip,trim= 20 0 0 0,width=0.8\linewidth,
       angle=0]{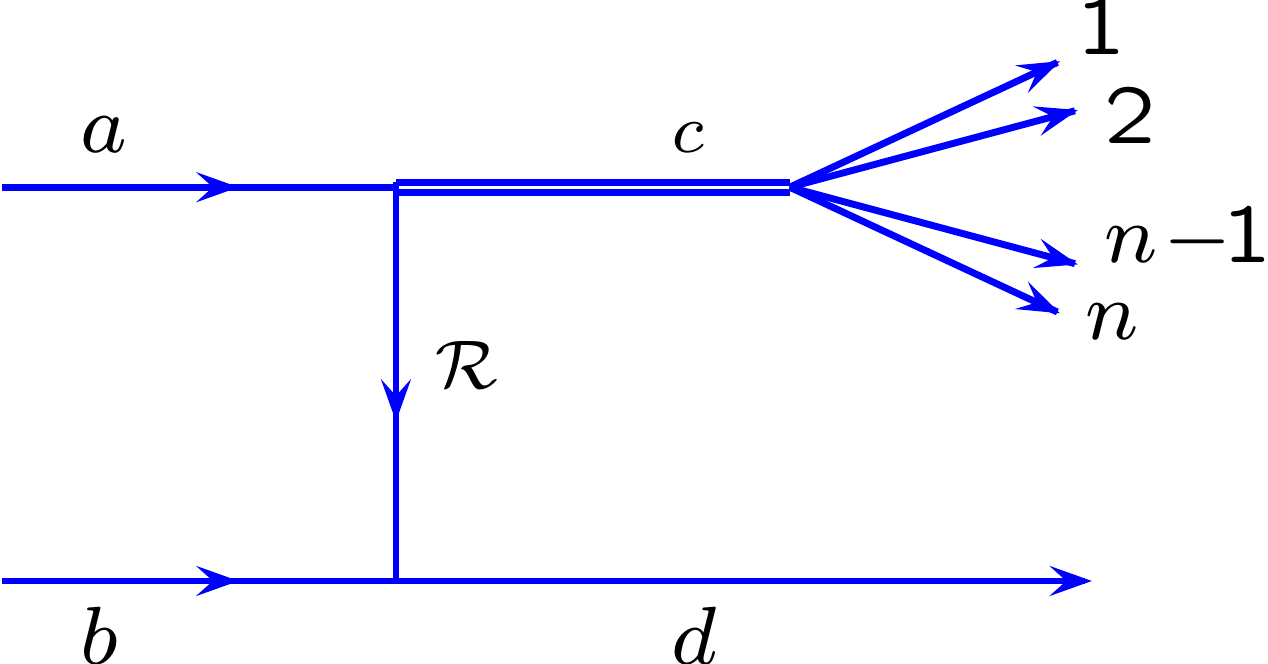}
\vspace{0.3cm}
      \includegraphics[clip,trim= 17 10 3 -10,width=0.8\linewidth,
     angle=0]{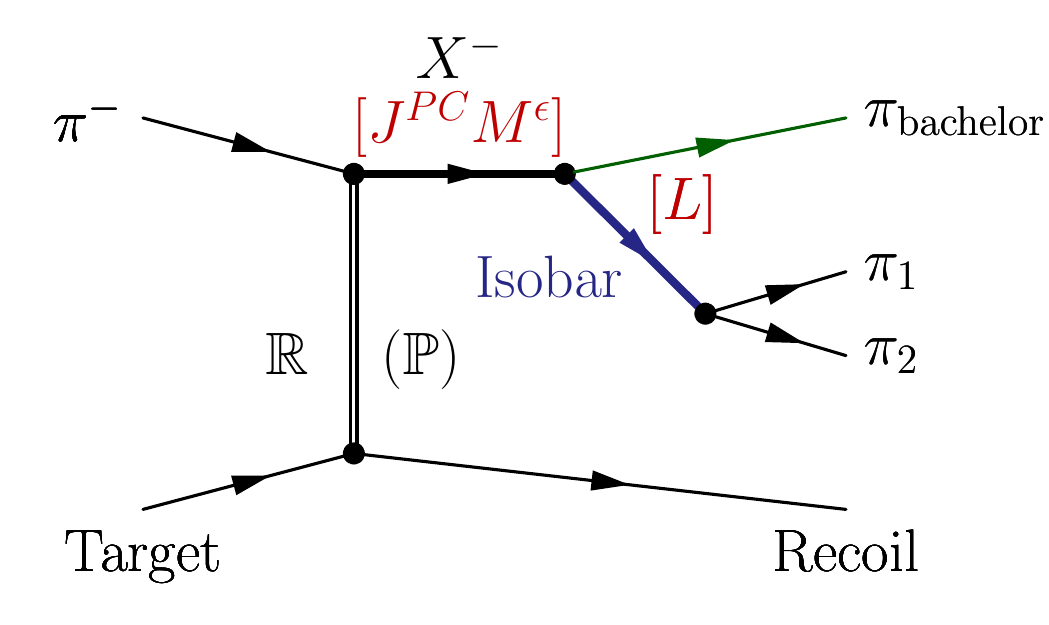}
    \end{center}
  \end{minipage}
  \hfill
  \begin{minipage}[h]{.59\textwidth}
    \begin{center}
     \includegraphics[clip, trim= 75 50 105 110,width=0.9\linewidth]{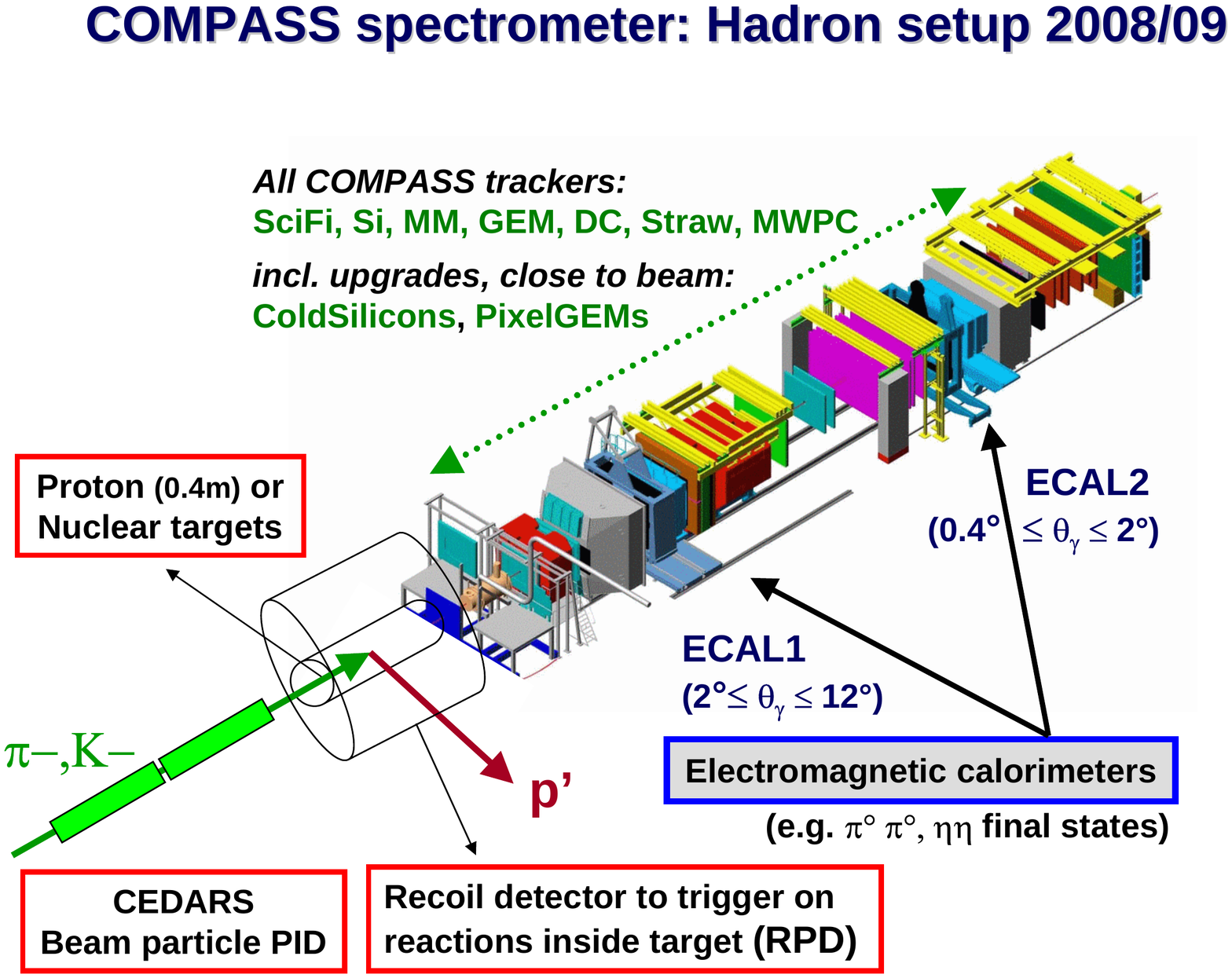}
      \caption{Left: \textit{(top)} Meson production in diffractive scattering via t-channel Reggeon exchange.
      \textit{(bottom)} Diffractive dissociation into 3$\pi$ final states as described in the isobar model: The diffractively produced resonance 
      $X^{-}$  with quantum numbers $J^{PC}M^\epsilon$ decays into an isobar with spin $S$ and relative orbital angular momentum $L$
      with respect to the $\pi_{\rm bachelor}$, the isobar subsequently decays into two pions.
      At high energies, the Pomeron is the dominant Regge-trajectory.
      Right: Sketch of the two-stage COMPASS spectrometer ($\sim$ 50\,m long) as used during hadron runs 2008 and 2009.}
      \label{fig:diffrProd_Spectro} 
    \end{center}
  \end{minipage}
\end{figure}

In \textit{diffractive pion dissociation} (at high energy), see Fig.~\ref{fig:diffrProd_Spectro} (left/top), the incident beam particle $a$ is excited via (t-channel) 
Reggeon exchange to some resonance $c$, which further dissociates into $n$ final state particles, whereas the target particle $b$ remains intact: 
$a+b \rightarrow c + d,~{\rm with}~ c \rightarrow 1+...+n$ particles, and $d$ denotes the recoil (target) particle. 
Interactions of this type are characterised by two kinematic variables $s$ and $t'=|t|-|t|_{\rm min}$, where $s=(p_{\rm a}+p_{\rm b})^2$ is the 
squared centre-of-mass energy and $t=(p_{\rm a}-p_{\rm c})^2$ is the square of the four momentum transfered from the incident beam particle to the outgoing system $c$.
Depending on the produced invariant mass $m_{\rm c}$, a minimum value of $|t|_{\rm min}$ is allowed by kinematics, which is small but larger than zero due to 
the longitudinal four-momentum transfer needed ($m_{\rm c} > m_{\rm a}$). In the centre-of-mass system: 
\begin{equation}
t'=|t|-|t|_{\rm min}=2|\vec p_{\rm a}||\vec p_{\rm c}|(1-\cos\theta_{\rm 0}) \ge 0 
~~~ {\rm with} ~~~ 
|t|_{\rm min} = 2(E_{\rm a}E_{\rm c} - |\vec p_a||\vec p_b|) - (m_{\rm a}^2 + m_{\rm c}^2)  ~~,
\label{Eq:T_prime}
\end{equation}
where $\theta_{\rm 0}$ is the scattering angle. Diffractive reactions have a total cross section in the order of 1--2\,mb. Even though the 
differential cross section drops as $1/m_{\rm c}$, states beyond 3\,GeV/c$^2$ can be produced diffractively in a fixed target experiment like 
COMPASS (190\,GeV/c$^2$ $\pi$ beam, proton target). Due to the forward kinematics the final state particles have to be detected mostly under small angles 
(with respect to the beam) requiring excellent angular resolution.
   
\section{Experimental set-up in 2008/09}
A detailed description of the COMPASS two-stage spectrometer (Fig.\,\ref{fig:diffrProd_Spectro} (right)) dedicated to a variety 
of fixed-target physics programmes can be found in \cite{compass:2007}. 
For the measurement with hadron beams started in 2008, a 40\,cm long liquid hydrogen target with a diameter of 35\,mm, 
or simple disks of solid material (part of 2009 run) have been used. 
%
\begin{figure}[tp!]
  \begin{minipage}[h]{.335\textwidth}
    \begin{center}
      \includegraphics[clip,trim= 5 15 20 43,width=0.9\linewidth,
       angle=90]{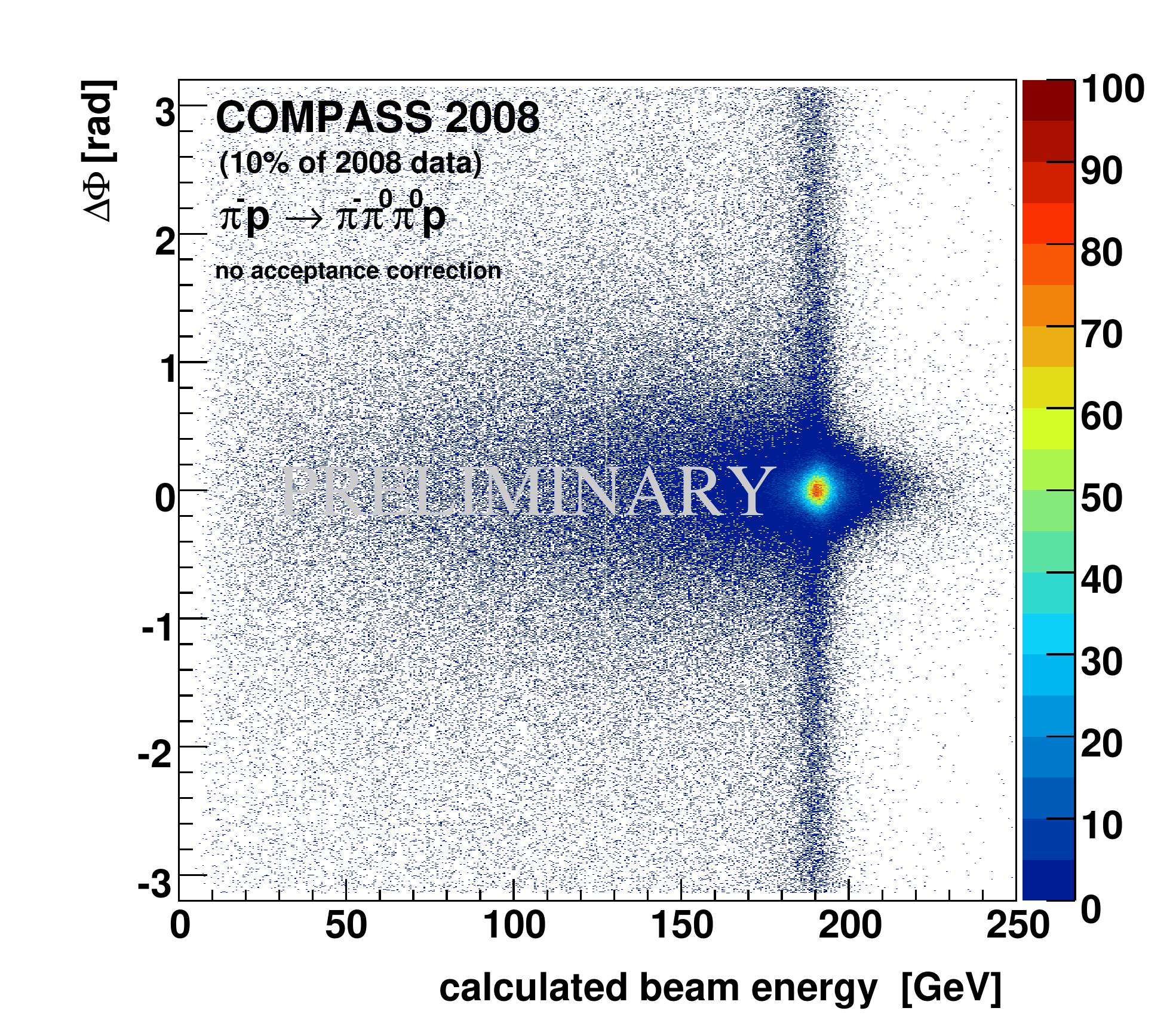}
    \end{center}
  \end{minipage}
  \hfill
  \begin{minipage}[h]{.33\textwidth}
    \begin{center}
      \includegraphics[clip,trim= 5 0 30 10,width=0.9\linewidth,
       angle=90]{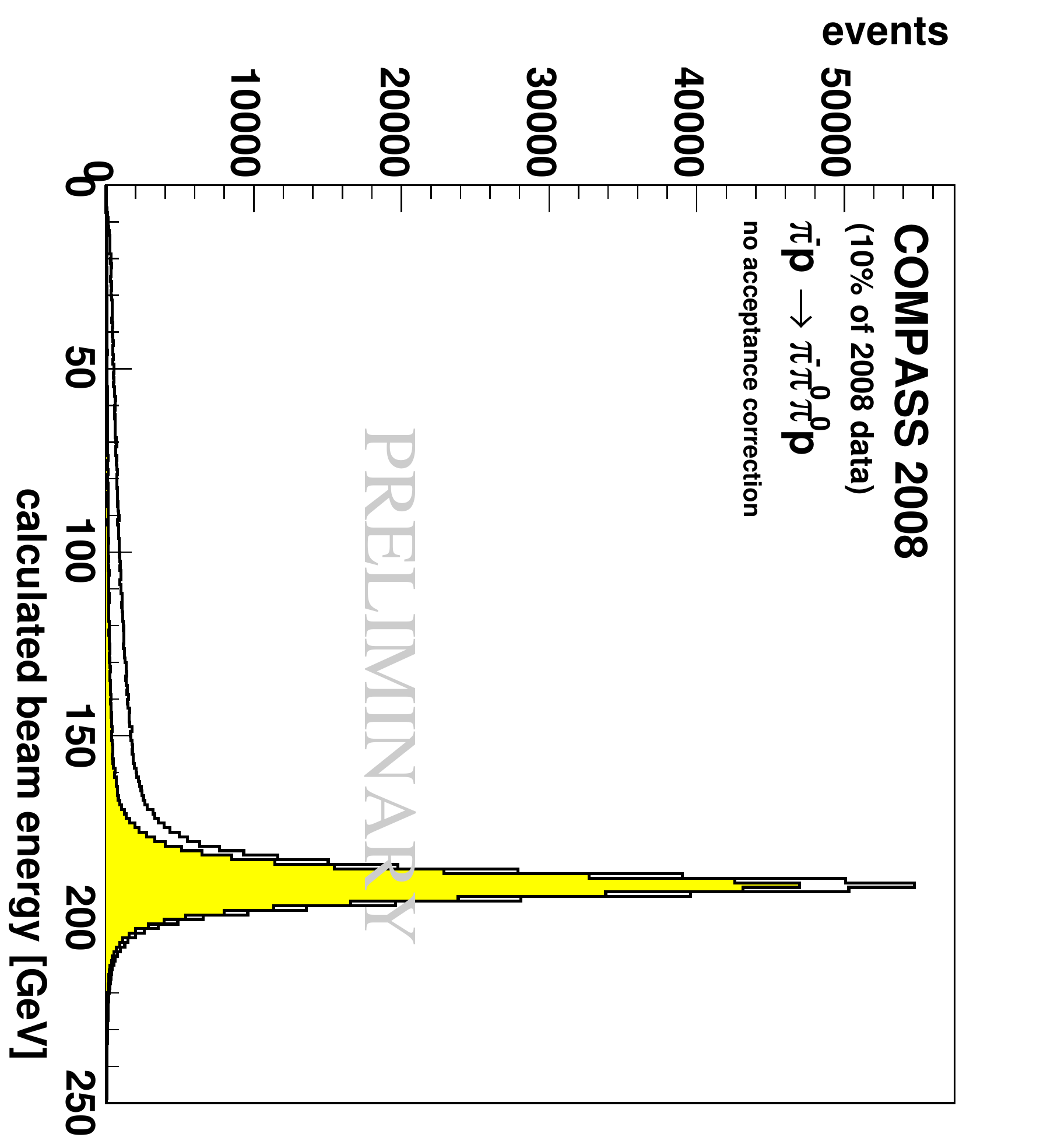}
    \end{center}
  \end{minipage}
  \hfill
  \begin{minipage}[h]{.335\textwidth}
    \begin{center}
      \includegraphics[clip,trim= 2 20 28 20,width=0.86\linewidth,
     angle=90]{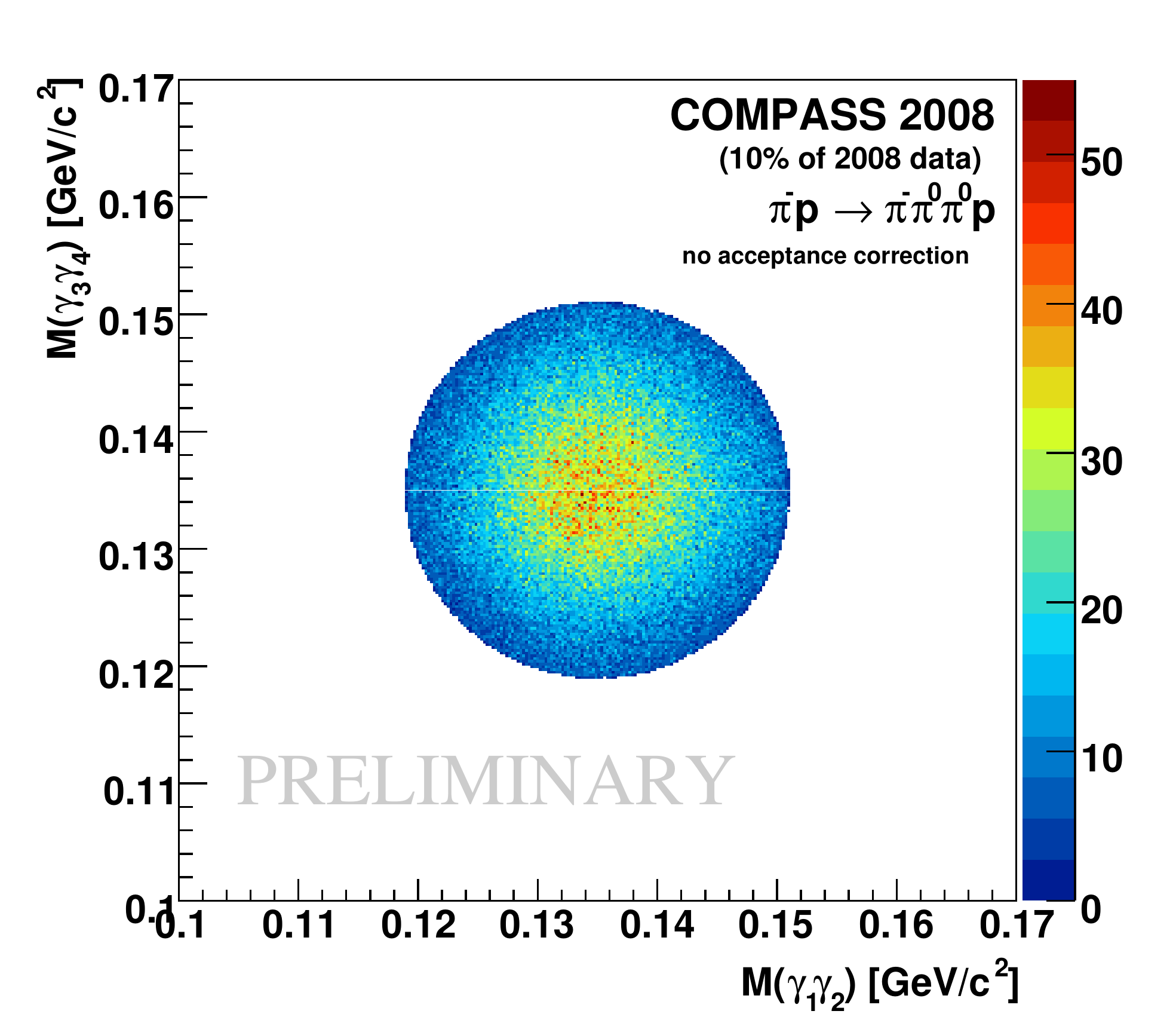}
      \caption{Exclusive events are selected by three mains cuts: $\Delta\Phi$, exclusivity, and $\pi^{0}$ mass. Left: $\Delta\Phi$ vs.
      calculated beam energy (i.e. exclusivity). Centre: Exclusivity before/after $\Delta\Phi$ cut. Right: Invariant mass of
      $\gamma_{\rm 1}\gamma_{\rm 2}$ vs. $\gamma_{\rm 3}\gamma_{\rm 4}$, cf. discussion in text.}
      \label{fig:EvtSelectionA}
    \end{center}
  \end{minipage}
\end{figure}
The spectrometer features electromagnetic and hadronic calorimetry in both stages. Photon 
detection in a wide angular range with high resolution is crucial for decay channels involving $\pi^{0}$, $\eta$ or $\eta'$.
Therefore, the read-out electronics have been upgraded (from 10 to 12\,bit SADCs) in 2008, allowing for Digital Signal Processing 
of the ADC signals, and 800 of the lead glass Cherenkov counters (3000 in total) have been replaced by so-called Shashlik sampling 
calorimeters in the central part of ECAL2 to improve the radiation hardness as well as the energy resolution, see e.g. 
\cite{fnerling:2008}. A new monitoring laser system for improved gain control of ECAL1 has further been installed in 2009. 
A Recoil Proton Detector (RPD) consisting of 2 concentric barrels of scintillator slats read out by PMTs was introduced to trigger 
on interactions inside the target and to detect the recoil particle. It performs a time-of-flight measurement at high 
accuracy ($\sim$~350\,ps). Finally, two CEDAR were installed, to separate the kaons ($\sim$ 2.5\,\%) in the beam from the pions 
(or, in case of proton beam, the pion contribution from protons).
\begin{figure}[bp!]
  \begin{minipage}[h]{.33\textwidth}
    \begin{center}
      \label{fig:exclusivity}
      \includegraphics[clip,trim= 5 25 25 7,width=0.91\linewidth,
     angle=90]{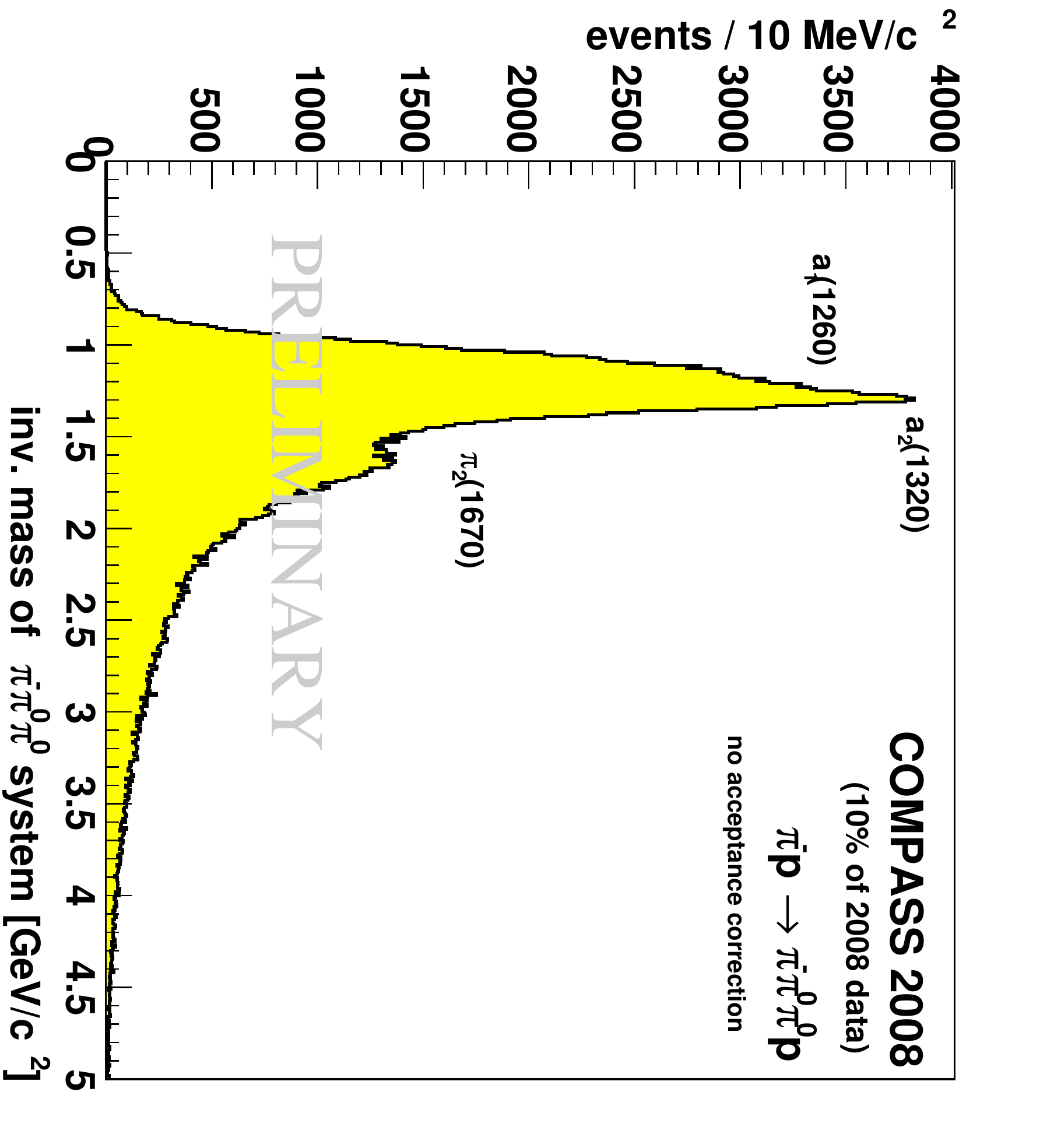}
    \end{center}
  \end{minipage}
  \hfill
  \begin{minipage}[h]{.33\textwidth}
    \begin{center}
      \includegraphics[clip,trim= 10 22 20 10,width=0.91\linewidth,
       angle=90]{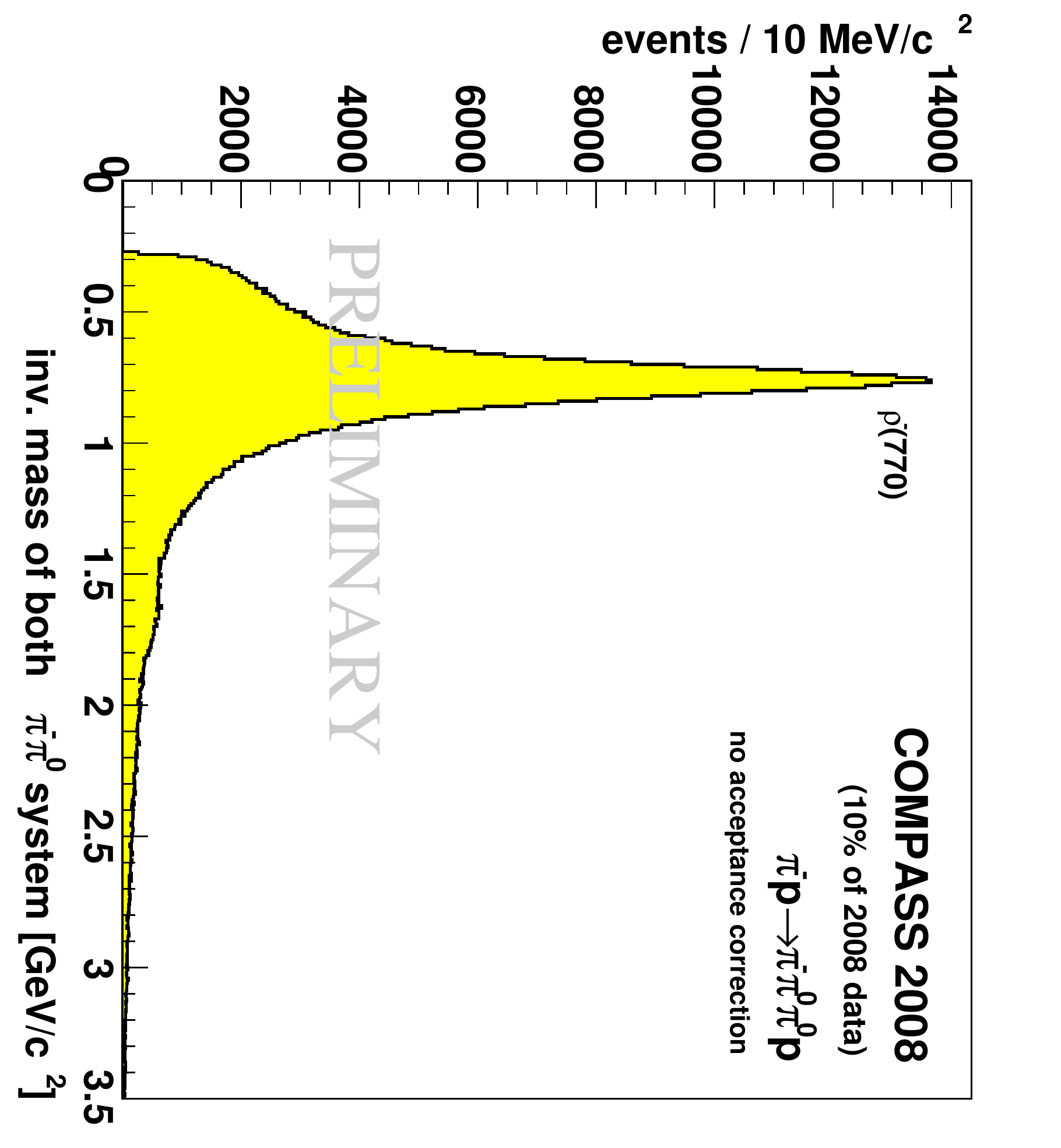}
    \end{center}
  \end{minipage}
  \hfill
  \begin{minipage}[h]{.33\textwidth}
    \begin{center}
      \includegraphics[clip,trim= 3 20 19 10,width=0.91\linewidth,
     angle=90]{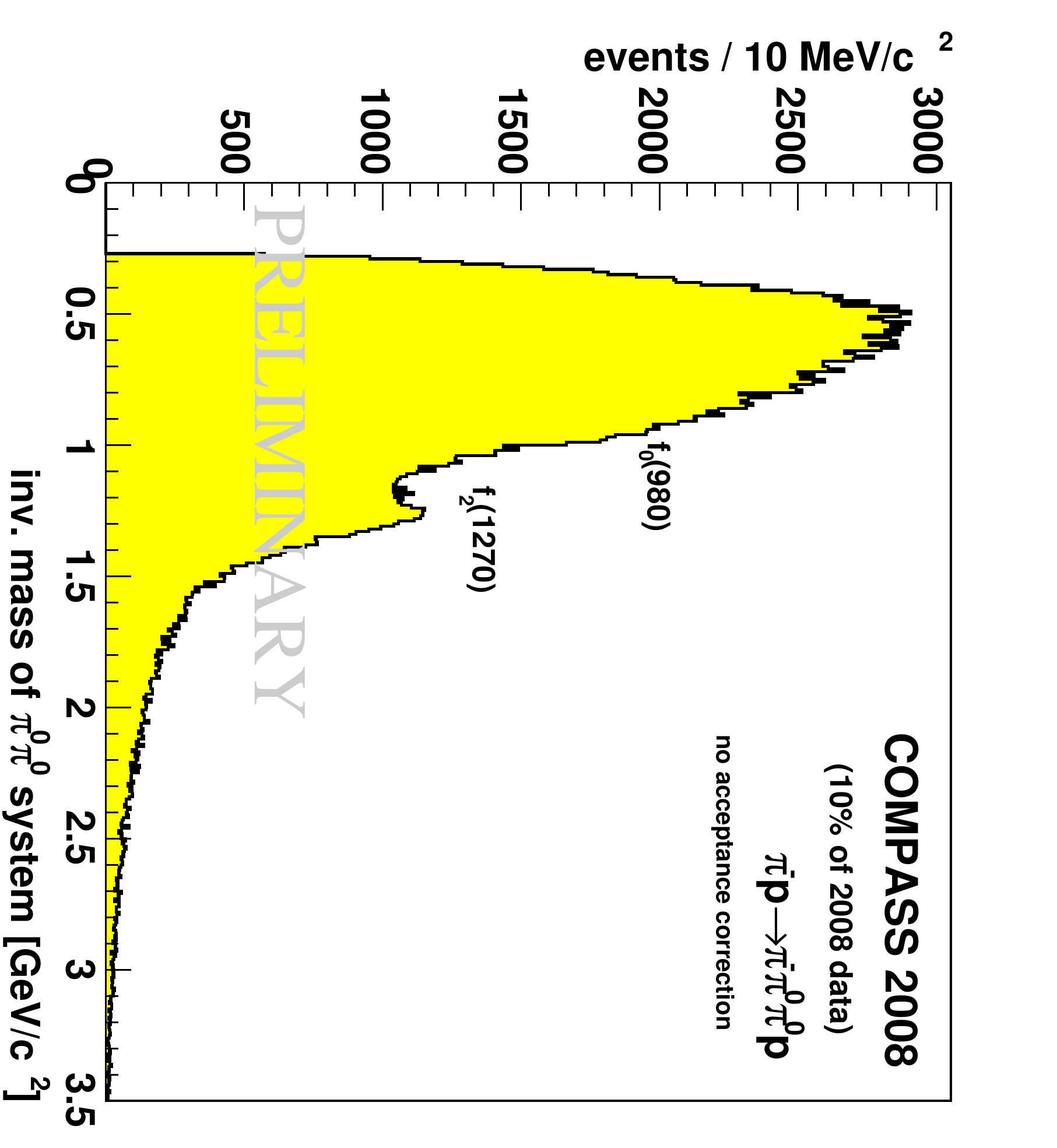}
      \caption{Invariant mass spectra of -- Left: Total outgoing system. Centre: $\pi^{-}\pi^0$ system. Right: $\pi^{0}\pi^0$ system.}
      \label{fig:invMass}
    \end{center}
  \end{minipage}
\end{figure}
\section{Data selection}
The data presently analysed for diffractively produced $\pi^{-}\pi^{0}\pi^{0}$ final states corresponds to 
$\sim$ 10\,\% of the data taken with pion beam in 2008. 
The diffractive trigger selected events with one incoming charged beam particle and a recoil proton detected by the RPD. 
Non-interacting beam and events out of acceptance were vetoed. Exactly one primary vertex inside the target volume is required 
for each event. Events with exactly one outgoing charged track and 4 $\gamma$s, from the two $\pi^{0}$ decays, 
detected in ECAL1 and ECAL2 were selected, if they give exactly one $\pi^{0}\pi^{0}$ combination within 
a circular cut of $\pm$ 20\,MeV/c$^2$ around the PDG mass (preselection, later tightened). Background events from elastic scattering 
have been suppressed by cutting on the energy $E_{\pi^{-}}< 185$\,GeV. 
In order to select exclusive events, three main cuts are applied consistently in terms of $\pm 2\,\sigma$ of each distribution of the 
three observables after having applied the other two.
\begin{figure}[tp!]
  \begin{minipage}[h]{.32\textwidth}
    \begin{center}
      \label{fig:exclusivity}
      \includegraphics[clip,trim= 3 30 42 10,width=0.9\linewidth,
       angle=90]{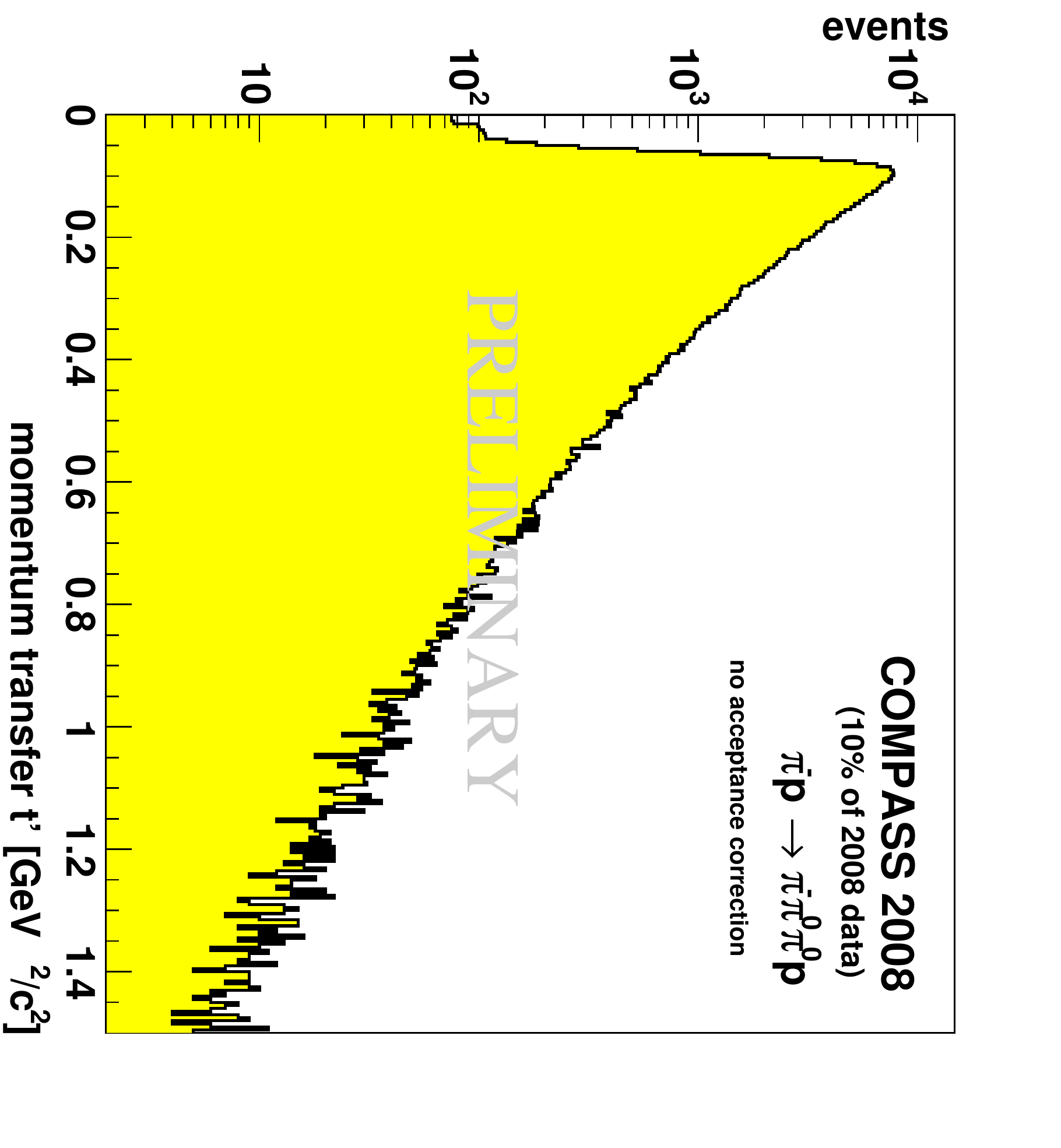}
    \end{center}
  \end{minipage}
  \hfill
  \begin{minipage}[h]{.33\textwidth}
    \begin{center}
      \includegraphics[clip,trim= 3 12 42 25,width=0.86\linewidth,
       angle=90]{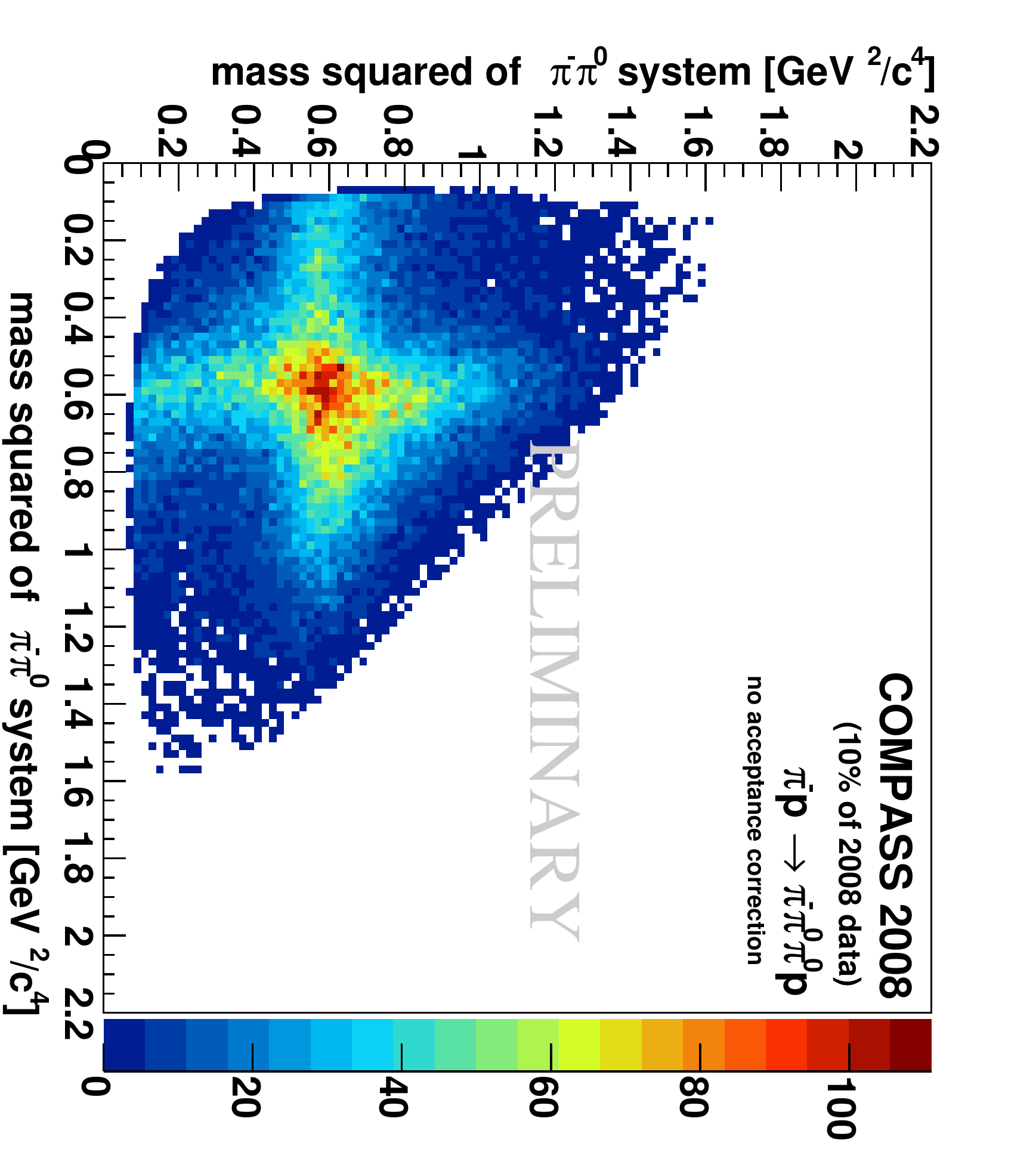}
    \end{center}
  \end{minipage}
  \hfill
  \begin{minipage}[h]{.35\textwidth}
    \begin{center}
      \includegraphics[clip,trim= 3 23 42 22,width=0.82\linewidth,
       angle=90]{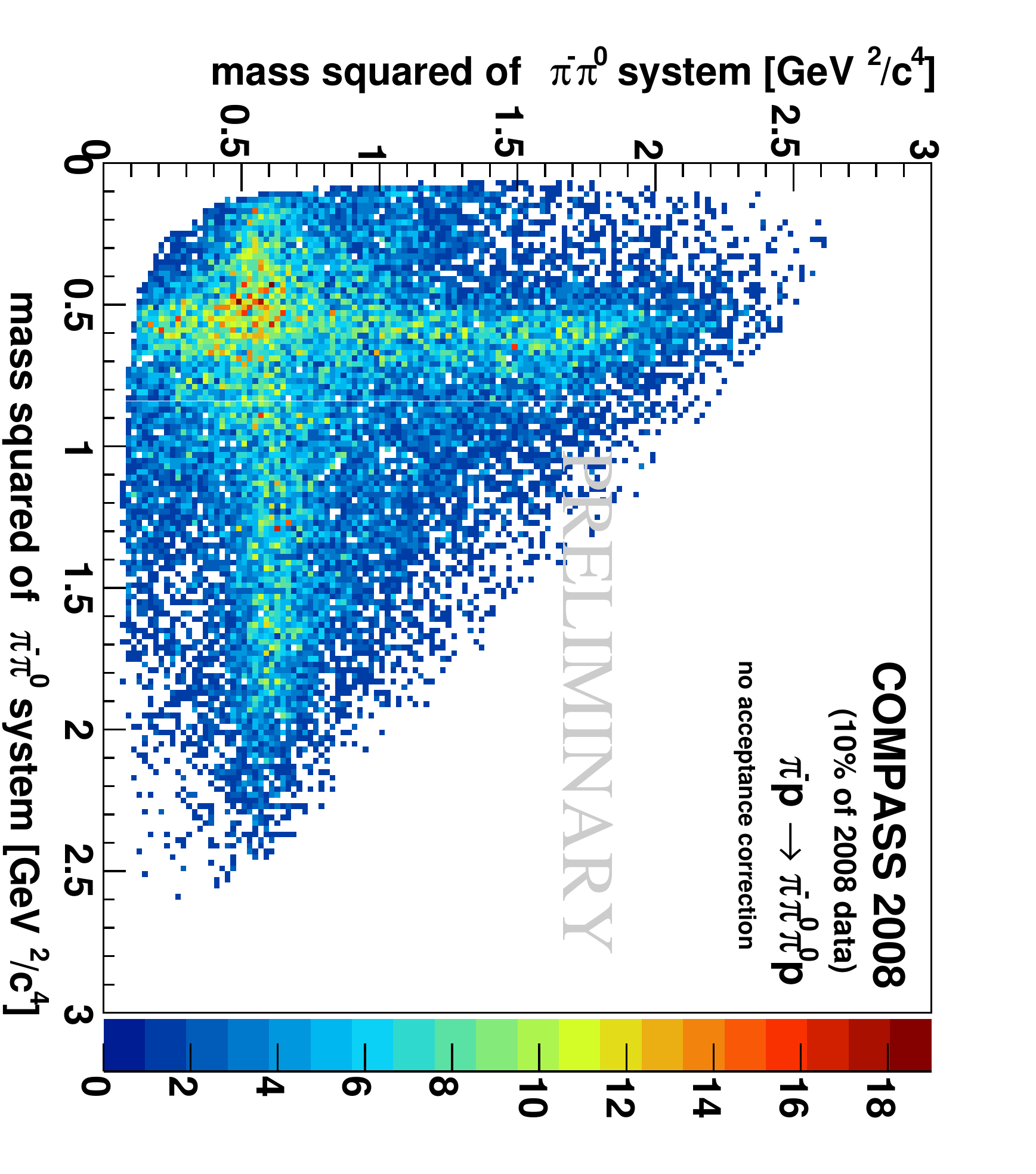}
      \caption{Left: Distribution of squared four-momentum transfer $t'$ (Eq.\ref{Eq:T_prime}). 
               Centre: Dalitz plot in $a_{\rm 2}$ region ($1.320\pm 0.100$\,GeV/c$^2$). 
               Right: Dalitz plot in $\pi_{\rm 2}$ region ($1.670\pm 0.100$\,GeV/c$^2$).}
      \label{fig:EvtSelectionB}
    \end{center}
  \end{minipage}
\end{figure}
Those are the angle $\Delta \Phi$  ($\pm 0.2$\,rad), defined as the azimuthal angle between the total momentum of 
the outgoing pion system measured with the spectrometer and the one of the recoil proton detected with the RPD, which should be
anti-parallel by momentum conservation, the exclusivity ($\pm 6$\,GeV) applied on the beam energy calculated from the 
outgoing system under assumption of energy conservation, and the $\pi^{0}$ mass ($\pm$16\,MeV). The corresponding distributions 
are shown in Fig.\,\ref{fig:EvtSelectionA} (left), the exclusive sample appears around the nominal beam energy. 
The background below the exclusivity peak is suppressed by applying the cut on $\Delta \Phi$ as clearly seen in
Fig.\,\ref{fig:EvtSelectionA} (centre). The final $\gamma_1\gamma_2$ versus $\gamma_3\gamma_4$ invariant mass distribution after all 
three 2\,$\sigma$ cuts is given by Fig.\,\ref{fig:EvtSelectionA} (right).
It should be noticed that demanding exactly 4 ECAL clusters reduces presently the statistics outcome significantly, however, 
this will improve once our electromagnetic reconstruction is finalised, taking full advantage of the detector upgrade.
The resultant invariant mass distributions are given in Fig.\,\ref{fig:invMass}: The total 3$\pi$ system (left) looks similar to the charged
mode \cite{fhaas:2009}, clearly visible are the prominent $a_{\rm 1}(1260)$, $a_{\rm 2}(1320)$ and $\pi_{\rm 2}(1670)$.
The $\rho^{-}$ is cleanly seen in the $\pi^{-}\pi^{0}$ mass spectrum (centre) as well as the $f_{\rm 2}(1270)$ in $\pi^{-}\pi^{0}$ 
(right), also we might see the $f_{\rm 0}(980)$ in the $\pi^{0}\pi^{0}$ spectrum. Fig.\,\ref{fig:EvtSelectionB} (left) shows 
the $t'$ distribution for the final sample of 240\,k events. In Fig.\ref{fig:EvtSelectionB} (centre/right) 
the Dalitz plots in the $a_{\rm 2}$ and $\pi_{\rm 2}$ region, respectively, are shown. The main decays into the $\rho^{-}(770)$ are seen including the 
effect of constructive interference. Otherwise for the $\pi_{\rm 2}$ region, the $f_{\rm 2}(1270)$ is not yet visible due to lack of statistics.    

\section{Partial Wave Analysis (mass independent)}
A PWA has been performed restricted to the range $0.1 \le t'\le 1.0$, to stay above the RPD threshold and to ensure diffractive reactions. 
In order to determine all resonances present in the data, including the quantum numbers, we perform our PWA in two steps: a mass independent PWA and a subsequent mass dependent fit. 
The program used was originally developed in Illinois and modified at Protvino and Munich \cite{jhansen:1973,dima}. 
At this first glance, essentially the same model that was used to analyse the 2004 data \cite{Alekseev:2009a} is applied to the 2008 
data to the neutral and charged \cite{fhaas:2009} decay modes started to be analysed. Five isobars, the $\rho(770)$, $f_{\rm 0}(980)$, $f_{\rm 2}(1270)$, $\rho_{\rm 3}(1690)$, and 
the $(\pi\pi)_{\rm s}$, as a parameterisation of the broad $\sigma(600)$ and $f_{\rm 0}(1370)$ \cite{ikachaev:2001}, are included. 
In total 42 partial waves are fitted to the data, including a background wave, which is flat in the relevant Gottfried-Jackson (GJ) 
angles and added incoherently. We discuss the first preliminary mass-independent fits, where the angular distributions are fitted in 
40\,MeV/$c^2$ bins of the 3$\pi$ invariant mass $m=m_{\rm c}$, assuming the production strength for a given wave to be constant within a given mass bin. 
\begin{figure}[tp!]
  \begin{minipage}[h]{.49\textwidth}
    \begin{center}
      \includegraphics[clip,trim= 3 4 22 5,width=0.9\linewidth,
       angle=0]{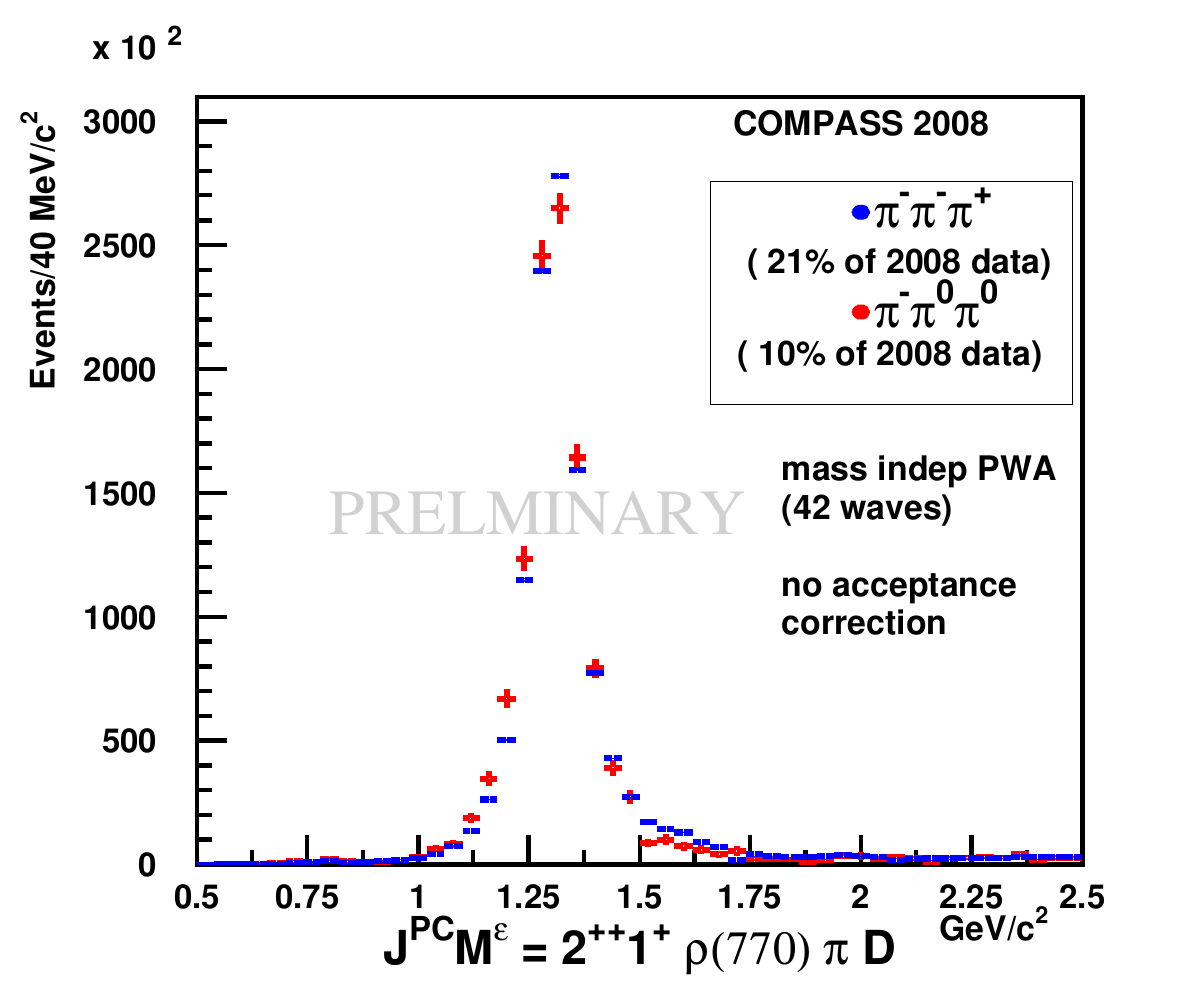}
    \end{center}
  \end{minipage}
  \hfill
  \begin{minipage}[h]{.49\textwidth}
    \begin{center}
      \includegraphics[clip,trim= 3 4 22 5,width=0.9\linewidth,
     angle=0]{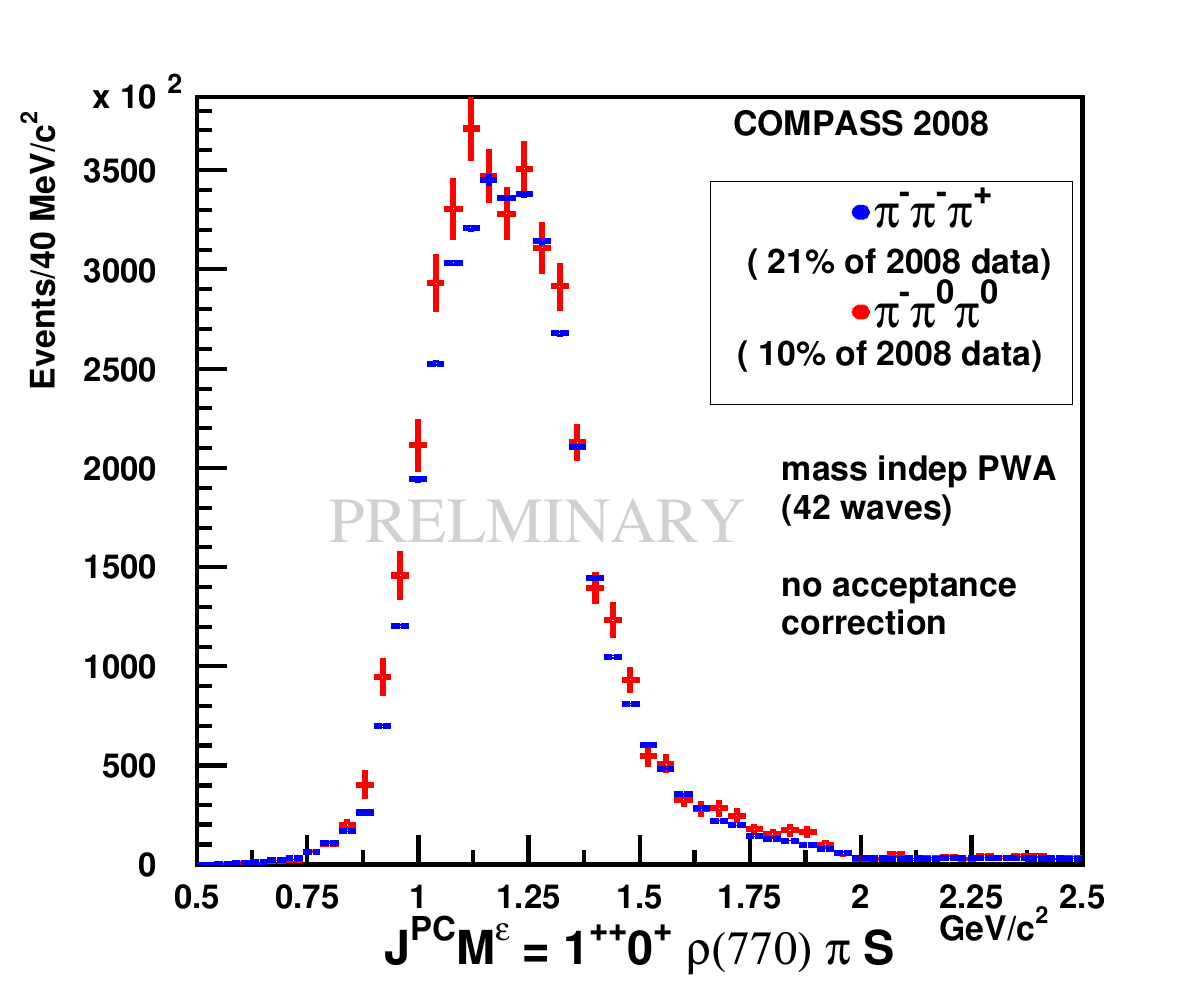}
      \caption{Comparison of PWA intensities of main waves for neutral vs. charged mode. 
       Left: Intensities of the $a_{\rm 2}$ ($2^{\rm ++}1^{\rm +}$ going into $\rho^{-} \pi$ D wave) used for 
	     normalisation of charged to neutral mode.      
       Right: ($a_{\rm 1}$) $1^{\rm ++} 0^{\rm +}$ into $\rho^{-} \pi$ D wave. }
      \label{fig:PWA}
    \end{center}
  \end{minipage}
\end{figure}
The underlying formalism is based on two assumptions. The total cross-section is separated into a resonance and a recoil vertex, 
and the isobar model, see Fig.\,\ref{fig:diffrProd_Spectro} (left/bottom), is used to describe the $X^{-}$ decay into three pions 
as a two-step process, without any further final state interactions among the pions nor with the target.
The decay goes via an intermediate $I=0$ di-pion resonance, the so-called isobar, decaying into a pion pair, and a 
so-called bachelor pion. The isobar spin $S$ and relative orbital angular momentum $L$ between the isobar and the bachelor pion 
couple to the spin $J$ of the resonance $X^{-}$. 
The 3$\pi$ system has isospin $I>0$ in general, and we can assume $I=1$ as no flavour-exotic mesons are known in the light quark sector. 
Since the final state under study comprises an odd number of pions (and thus negative G-parity), the charge conjugation is positive. 
The amplitudes are constructed in the reflectivity basis \cite{schung:1974} so that the $X^{-}$ spin projection $M \ge 0$ and the 
reflectivity $\epsilon = \pm 1$ describes the symmetry under reflection at the production plane. Amplitudes of different
reflectivities do not interfere due to parity conservation. Moreover, at high $\sqrt{s}$ the reflectivity equals naturality 
of the exchanged Reggeon, $\epsilon=+1$ corresponds to natural parity exchange like e.g. Pomeron mediated reactions. 
The full set of quantum numbers $J^{PC} M^{\epsilon}[isobar~\pi]L$ defines a partial wave, whereas $I$ and $G$ are not explicitly
specified since they are fixed by the incoming pion to $I^G=1^-$.
For the spin density matrix, we allow for rank $N_{\rm r}$=2 to account for helicity flip and non-flip amplitudes at the baryon vertex, 
assuming the target nucleon stays intact. Finally, the observed intensities are parameterised as a coherent and incoherent sum over the 
partial wave amplitudes \cite{schung:1974}:
\begin{equation}
\sigma_{\rm indep}(\tau,m) = \sum_{\epsilon=\pm 1}\sum_{\rm r = 1}^{N_{\rm r} } ~\biggr|\sum_{i}T^{\epsilon}_{ir} \Psi^{\epsilon}_{i}(\tau,m)
\biggr/ \sqrt{\int |\Psi^{\epsilon}_{i}(\tau',m)|^2{\rm d}\tau'}~\biggr|^2~~, 
\end{equation}
where the three body kinematics are described completely by five phase space coordinates represented by $\tau$, measured for each event. 
They are the input for calculating the decay amplitudes, $\Psi^{\epsilon}_{i}$ for each partial wave $i$, using the D-function formalism
in the helicity frame. 
The complex numbers $T^{\epsilon}_{ir}$, the so-called production amplitudes, contain information of strength and interference of the 
waves. They are obtained using an extended maximum-likelihood method. The spectrometer acceptance can be taken into account directly in this procedure. 
 
\section{First results, check of isospin symmetry}
\begin{figure}[tp!]
  \begin{minipage}[h]{0.55\textwidth}
    \begin{center}
    \includegraphics[clip,trim= -10 15 55 30, width=0.95\linewidth, angle=0]{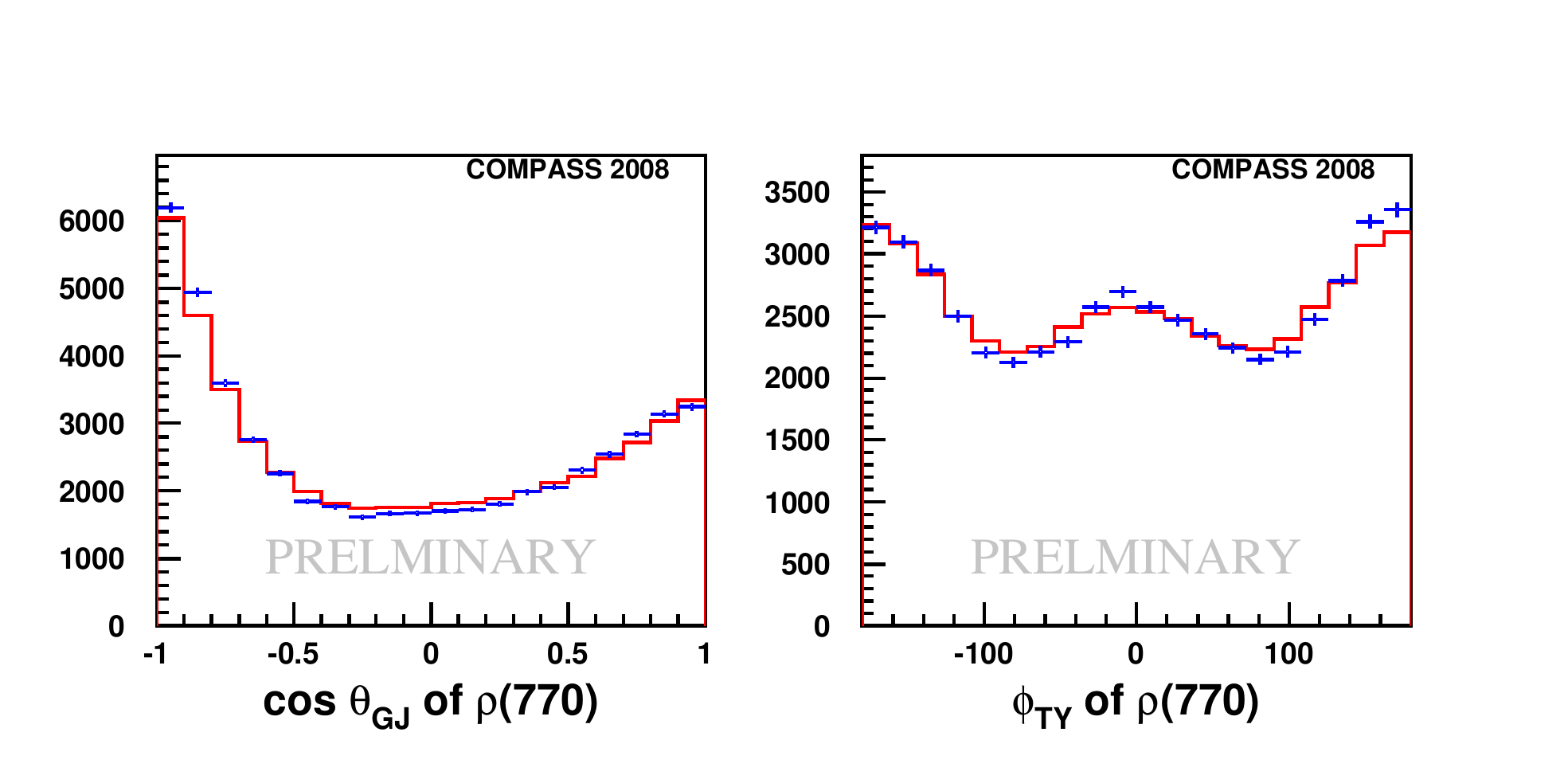}
    \end{center}
  \end{minipage}
  \begin{minipage}[h]{0.55\textwidth}
    \begin{center}
     \includegraphics[clip,trim= 20 20 25 35, width=0.95\linewidth, angle=0]{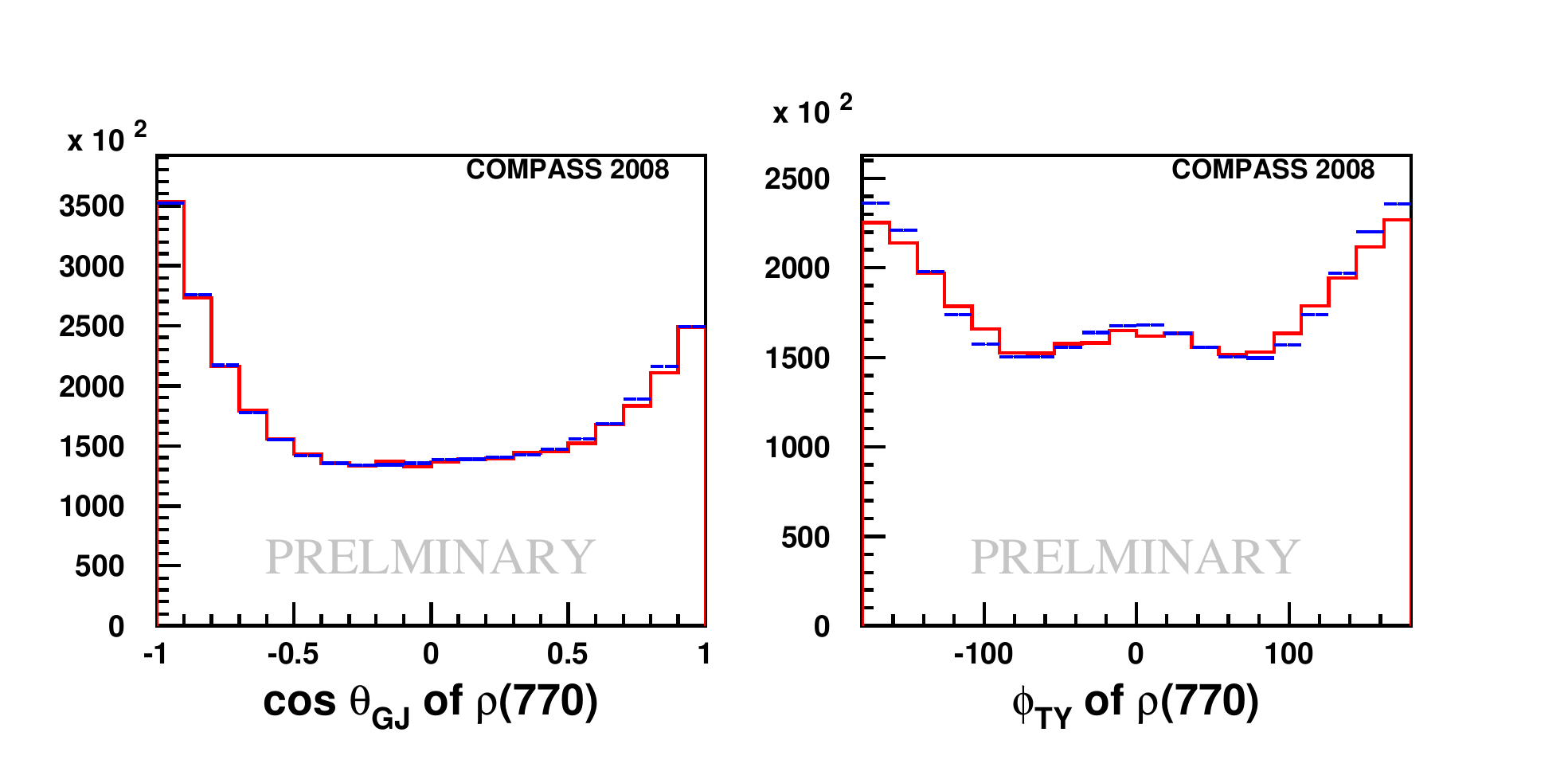}
    \caption{Comparison of neutral (left) and charged (right) mode data to a phase space MC weighted with production and decay 
amplitudes from the fit result, under assumption of a uniform acceptance: Direction of the $\rho(770)$, $\cos \theta$ and $\phi$, in the 
GJ frame - limited to the $a_1/a_2$ region.}
      \label{fig:pwa_predict}
    \end{center}
  \end{minipage}
\end{figure}

As mentioned before, simultaneous observation of both $(3\pi)^{-}$ modes in the same experiment provides important cross-check. When an
isospin 1 resonance $X^{-}$ is produced, and subsequently decays via an intermediate $I=0$ di-pion resonance, the yield in the neutral mode should be half of that in the charged mode. Otherwise, if the di-pion is an isovector, equal yields are expected for both modes. 
\begin{figure}[tp!]
  \begin{minipage}[h]{.49\textwidth}
    \begin{center}
      \includegraphics[clip,trim= 3 4 22 5,width=0.9\linewidth,
	angle=0]{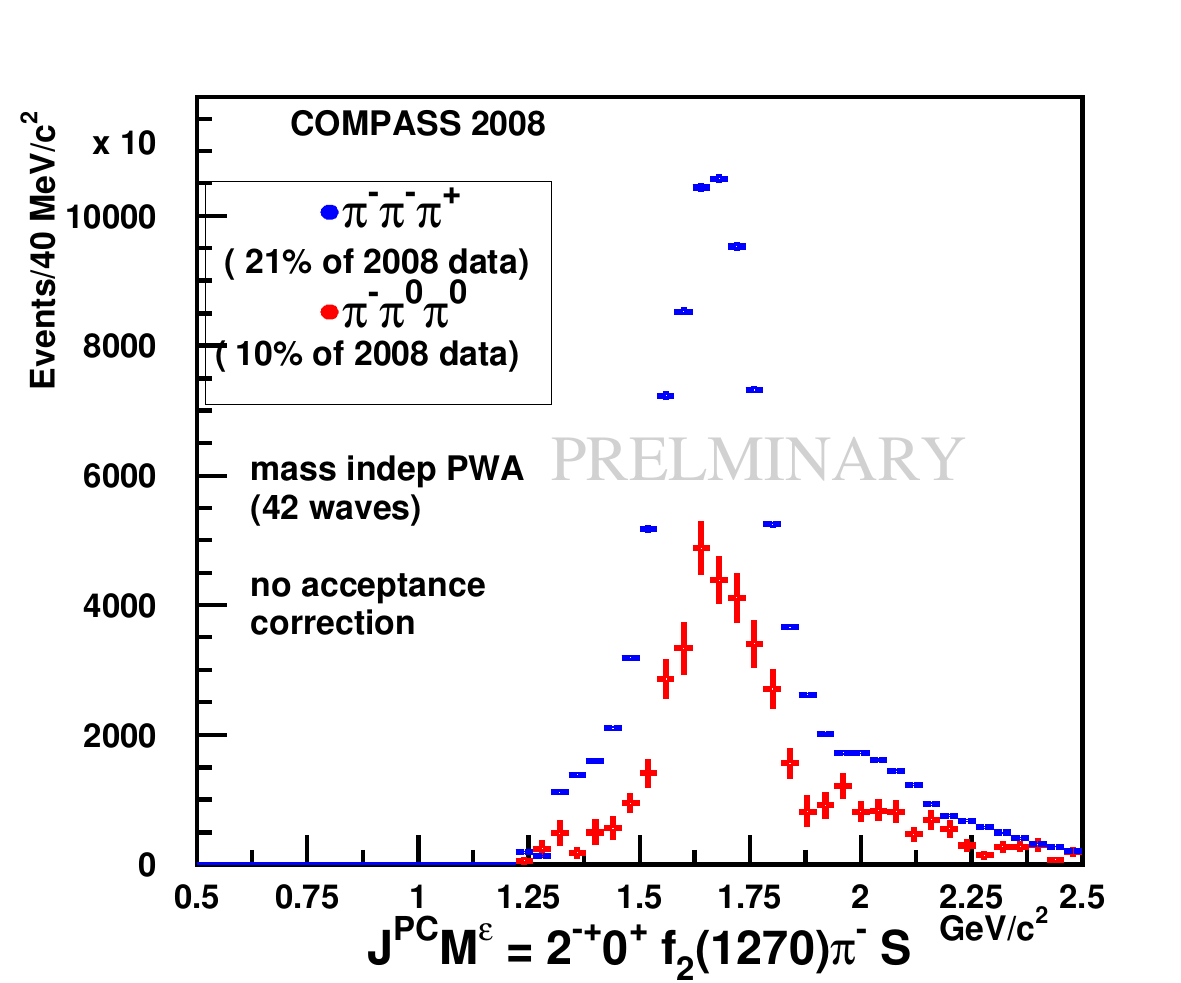}
    \end{center}
  \end{minipage}
  \hfill
  \begin{minipage}[h]{.49\textwidth}
    \begin{center}
      \includegraphics[clip,trim= 3 4 22 5,width=0.9\linewidth,
     angle=0]{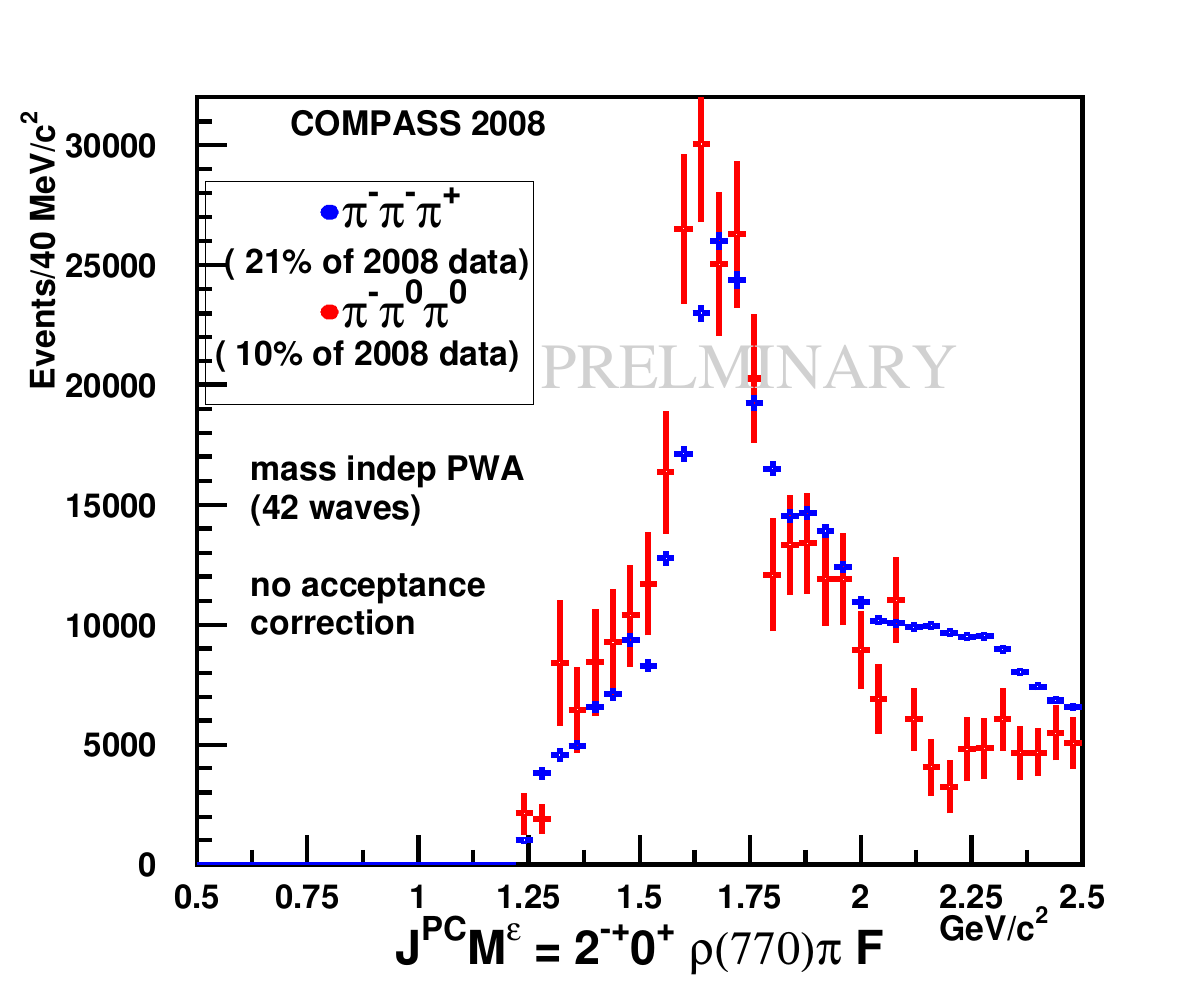}
      \caption{Comparison of PWA intensities of main waves for neutral vs. charged mode. 
      Left: ($\pi_{\rm 2}$) $2^{\rm -+} 0^{\rm +}$ into $f_{\rm 2}$(1270) $\pi$ S wave. 
      Right: ($\pi_{\rm 2}$) $2^{\rm -+} 0^{\rm +}$ into $\rho^{-} \pi$ F wave.}
      \label{fig:PWA_b}
    \end{center}
  \end{minipage}
\end{figure}
This isospin symmetry holds only if the branchings are completely determined by the Clebsch-Gordan coefficients. In general, 
however, Bose-Symmetrisation with the bachelor pion is obligatory and might affect the observed branchings. We checked by calculation (using
the wave functions) that the expectation for observed intensities as formulated before are correct for all isobars going to $\rho\pi$. 
Here, the effect might indeed not be negligible, but is the same for charged and neutral mode, and therefore cancels out. On the other hand,
the isospin symmetry needs to be modified for isobars going into $f_{0,2}\pi$ due to interference effects from Bose-Symmetrisation. For
example, in case of $0^{-+}$ into $f_{\rm 0}(1400)\pi~S$ wave, this effect doubles the expected suppression factor of two (simply expected
from Clebsch-Gordan coefficients) of intensities observed in the neutral versus the charged mode, see Tab.\ref{tab:isospinCheck}.
\begin{table}[bp]
  \centering
  \begin{tabular}[]{lll} \hline
    BR = N($\pi^-\pi^0\pi^0 $  )/N($\pi^-\pi^-\pi^+ $  ) -- calculated from isobar model amplitudes \\  \hline  
    BR( $ \rho \pi$) = 1.\\
    BR( $ 0^{-+} f_0(1400) \pi$~$  S $) = 0.26 (at 1.3 GeV) = 0.29 (at 1.8 GeV)\\
    BR( $ 0^{-+} f_0(980) \pi$~$ S $) =  0.44 (at 1.8 GeV) \\
    BR( $ 1^{++} f_0(1400) \pi $~$P $) = 0.80 (at 1.3 GeV) \\
    BR( $ 2^{-+} f_2(1270) \pi $~$ S $) =  0.50 (at 1.67 GeV) \\
    \hline 
  \end{tabular}
  \caption{Isospin symmetry checks: Calculation of branching ratios (BR) for charged and neutral mode and different 
       isobar decays. 
}
  \label{tab:isospinCheck}
\end{table}
In this extreme case, it is rather a factor of 4. Otherwise we find no distortion of the isospin symmetry due to such interference 
effects for the example of $2^{-+}$ into $f_{\rm 2}(1270)\pi~S$ wave (at the $\pi_{\rm 2}(1670)$ PDG mass), and expect here 
indeed to observe the pure suppression factor of two given by the Clebsch-Gordan coefficients. The calculated branching ratios summarised in Tab.\,\ref{tab:isospinCheck} are in good agreement with the data.
Even though acceptance has not yet been corrected for at this stage of the analysis, 
the dominant intensities show the symmetry as described above, and thus the acceptance is proved to be rather 
uniform. In order to compare the observed intensities in the neutral to the charged mode, the PWA results have been normalised to the
well-established narrow $a_{\rm 2}(1320)$ observed in both modes, see Fig.\,\ref{fig:PWA} (left).
Looking at $a_{1}$, $1^{++}0^{+}[\rho^{-}(770)\pi]~S$ wave (Fig.\,\ref{fig:PWA} (right)), we find the intensities as well as the widths being quite similar for the different modes, as expected. 
For the $\pi_{2}$, $2^{-+}0^{+}[f_{\rm 2}(1270)\pi]~S$ wave (Fig.\,\ref{fig:PWA_b} (left)), we obtain the neutral mode 
being suppressed by a factor $\sim 2.2$ relatively to the charged case, which is qualitatively already in good agreement with 
our expectation.
On the other hand looking at $\pi_{2}$, $2^{-+}0^{+}[\rho^{-}(770)\pi]~F$ wave (Fig.\,\ref{fig:PWA_b} (right)), we find consistently again about the same intensities.
As a quality check of the fits, we compare real data to Monte Carlo (MC) events, which were generated by weighting phase space events
with the production and decay amplitudes from the fit result under the assumption of a uniform acceptance. Such comparison for the decay 
angles ($\cos \theta$ and $\phi$) of the $\rho$ in the GJ frame are depicted for the neutral mode data in Fig.\,\ref{fig:pwa_predict} (left). The angles are shown for the limited mass region 
around $a_1$ and $a_2$ (1.22 to 1.38 GeV/$c^2$). Comparing the angles to the corresponding ones of the charged mode data, 
Fig.\,\ref{fig:pwa_predict} (right), limited to the same mass range around the $a_2$, one finds similar angular distributions, which is expected, since the physics is the same. Furthermore, for both cases the assumption of a flat acceptance seems to be valid, and therefore the comparison of both modes in terms of isospin symmetry is reasonable, even though the data has not yet been corrected for acceptance.
\vspace{-0.5cm}
\section{Summary \& outlook}
The COMPASS experiment has a high potential for contributing to light meson spectroscopy. Data with charged hadron beams on different targets with high statistics have been taken in 2008/09 with an upgraded apparatus. One main goal of the spectroscopy program is the 
search for $J^{PC}$ exotic states with gluonic degree of freedom, and to illuminate e.g. the disputed hybrid candidate $\pi_{\rm 1}(1600)$. 
In particular the detection of final states comprising both charged and neutral particles, allowing for cross-check and independent confirmation 
of any new state found, makes COMPASS unique as compared to previous fixed-target experiments.   
A first event selection and partial wave analysis of diffractively produced $\pi^{-}\pi^{0}\pi^{0}$ final states of a subset of the 2008 data (pion beam, 
proton target) has been performed. The observed main waves are, at this first glance, in good agreement with theoretical expectations 
(isospin symmetry), which demonstrates the feasibility of COMPASS for hadron spectroscopy not only of charged but also of channels involving 
neutral particles. The data recorded is of sufficient statistics to even study systematically the isobar model itself.  
Next steps for this analysis are increasing the statistics, application of acceptance corrections, extension of the waveset, and studying the existence of the exotic $1^{-+}$ wave in the 2008/09 data set.

\vspace{-0.5cm}
\begin{theacknowledgments}
\vspace{-0.2cm}
This work has been supported by the BMBF (Germany), 
particularly by the ``BMBF-Nutzungsinitiative CERN''.
\end{theacknowledgments}
\vspace{-0.1cm}


\bibliographystyle{aipproc}   
\vspace{-0.3cm}

\bibliography{nerling}

\IfFileExists{\jobname.bbl}{}
 {\typeout{}
  \typeout{******************************************}
  \typeout{** Please run "bibtex \jobname" to optain}
  \typeout{** the bibliography and then re-run LaTeX}
  \typeout{** twice to fix the references!}
  \typeout{******************************************}
  \typeout{}
 }

\end{document}